\documentstyle[12pt,psfig,epsf,isamu]{article}
\newcommand{\bet}{\begin{table}[h]}
\newcommand{\eet}{\end{table}}
\newcommand{\bef}[1]{\begin{figure}[#1]}
\newcommand{\eef}{\end{figure}}
\newcommand{\bec}{\begin{center}}
\newcommand{\eec}{\end{center}}
\newcommand{\ben}{\begin{enumerate}}
\newcommand{\een}{\end{enumerate}}
\newcommand{\bei}{\begin{itemize}}
\newcommand{\eei}{\end{itemize}}
\newcommand{\bed}{\begin{description}}
\newcommand{\eed}{\end{description}}
\newcommand{\beq}{\begin{equation}}
\newcommand{\eeq}{\end{equation}}
\newcommand{\bea}{\begin{eqnarray}}
\newcommand{\eea}{\end{eqnarray}}
\newcommand{\bey}{\begin{eqnarray*}}
\newcommand{\eey}{\end{eqnarray*}}

\def\p@enumi{$^\mbox{\arabic{enumi}}$}

\catcode`@=11 
\def\lsim{\mathrel{\mathpalette\@versim<}}
\def\gsim{\mathrel{\mathpalette\@versim>}}
\def\@versim#1#2{\lower0.2ex\vbox{\baselineskip\z@skip\lineskip\z@skip
  \lineskiplimit\z@\ialign{
          $\m@th#1\hfil##\hfil$\crcr#2\crcr\sim\crcr}}}

\begin{document}
\date{KEK-TH-467\\KEK Preprint 96-201\\OCHA-PP-72\\OU-HET 234\\
February 1996\\
}

\title{STUDYING THE HIGGS POTENTIAL VIA $e^+e^- \rightarrow Zhh$%
\footnote[4]{Talk presented by I.~Watanabe at the Workshop on 
Physics and Experiments with Linear Colliders (LCWS95), 
Morioka Appi Jappan, September 8-12, 1995.}
}

\author{Jun-ichi KAMOSHITA, Yasuhiro OKADA}
\address{Theory Group, KEK \\
         1-1 Oho, Tsukuba, Ibaraki 305, JAPAN \\
\small\rm E-mail: kamosita@theory.kek.jp \\
          E-mail: okaday@theory.kek.jp}
\author{Minoru TANAKA}
\address{Department of Physics, Osaka University \\
         1-16 Machikaneyama, Toyonaka, Osaka 560, JAPAN \\
\small\rm E-mail: minoru@theory.kek.jp}
\address{and}
\author{Isamu WATANABE}
\address{Department of Physics, Ochanomizu University \\
         2-1-1 Otsuka, Bunkyo, Tokyo 112, JAPAN \\
\small\rm E-mail: isamu@phys.ocha.ac.jp}

\maketitle
\abstracts{
  The physics prospect at future linear $e^+e^-$ colliders 
for the study of the Higgs triple self-coupling via
the process of $e^+e^-$ $\rightarrow Zhh$ is investigated.  
The measurement of this cross section leads us to the first 
non-trivial information on the Higgs potential. 

  We found that the Standard Model and the model without the Higgs 
self-coupling can be distinguished at the level of one standard 
deviation for a rather light Higgs mass with 100 fb$^{-1}$ integrated 
luminosity.  
  In the MSSM, the cross section is enhanced if  the production of 
at least one of heavy Higgs bosons ($H$ or $A$) and its 
subsequent decay ($H \rightarrow hh$, $A \rightarrow Zh$) are 
kinematically allowed.  
When such processes are not allowed, the cross section in the 
region of small $m_A$ is significantly  
suppressed relative to the SM cross section.
}

\newpage

\section{Introduction}

  In the Standard Model (SM) of particle physics, there are
three types of interactions of fundamental particles, 
gauge interactions, Yukawa interactions and the Higgs boson
self-interaction.  
It has already been confirmed by experiments that the
interactions between fermions and vector bosons are 
described by gauge theories.  
The gluon triple coupling has also been established.  
At LEP I$\!$I, LHC and  future linear colliders we will be able to 
examine the triple couplings of weak bosons and photons, the Higgs
couplings to $W^\pm$ and $Z$ bosons and the Yukawa couplings of 
heavy fermions.  

  To measure the Higgs self-coupling, the most promising way is
to observe the `Higgs double-production' processes, 
the processes with two Higgs bosons in the final state.
The cross sections of these processes in the SM have been evaluated
at $e^+e^-$ colliders%
$^{1-4}$, 
at hadron colliders%
$^{5-11}$
 and at $\gamma \gamma$ colliders%
$^{3,12}$.  
  The cross sections at linear colliders have been examined 
both in the SM and the model with an anomalous Higgs triple 
coupling%
$^{1-12}$.  
  Since the cross section is {\it relatively} large and all the 
final states can be identified without large missing momentum, 
the process $e^+e^-$ $\rightarrow Zhh$ is the best among 
the various Higgs double-production processes to look for the Higgs 
self-coupling during the first stage of a future linear collider. 

  In this paper we investigate the physics prospect of 
future linear $e^+e^-$ colliders for the study of the Higgs-boson
triple self-coupling via the process $e^+e^-$ $\rightarrow Zhh$, 
which leads us to the first non-trivial information 
on the Higgs potential. 
In Sec.\ 2 we examine the SM and a model without a Higgs
self-coupling, and we evaluate their distinguishability.  
Also, we survey the minimal supersymmetric standard model in 
Sec.\ 3.  
Conclusions will be given in Sec.\ 4.  

\bef{t}
\bec
\hspace*{1mm}
\epsfxsize=10cm
\epsffile{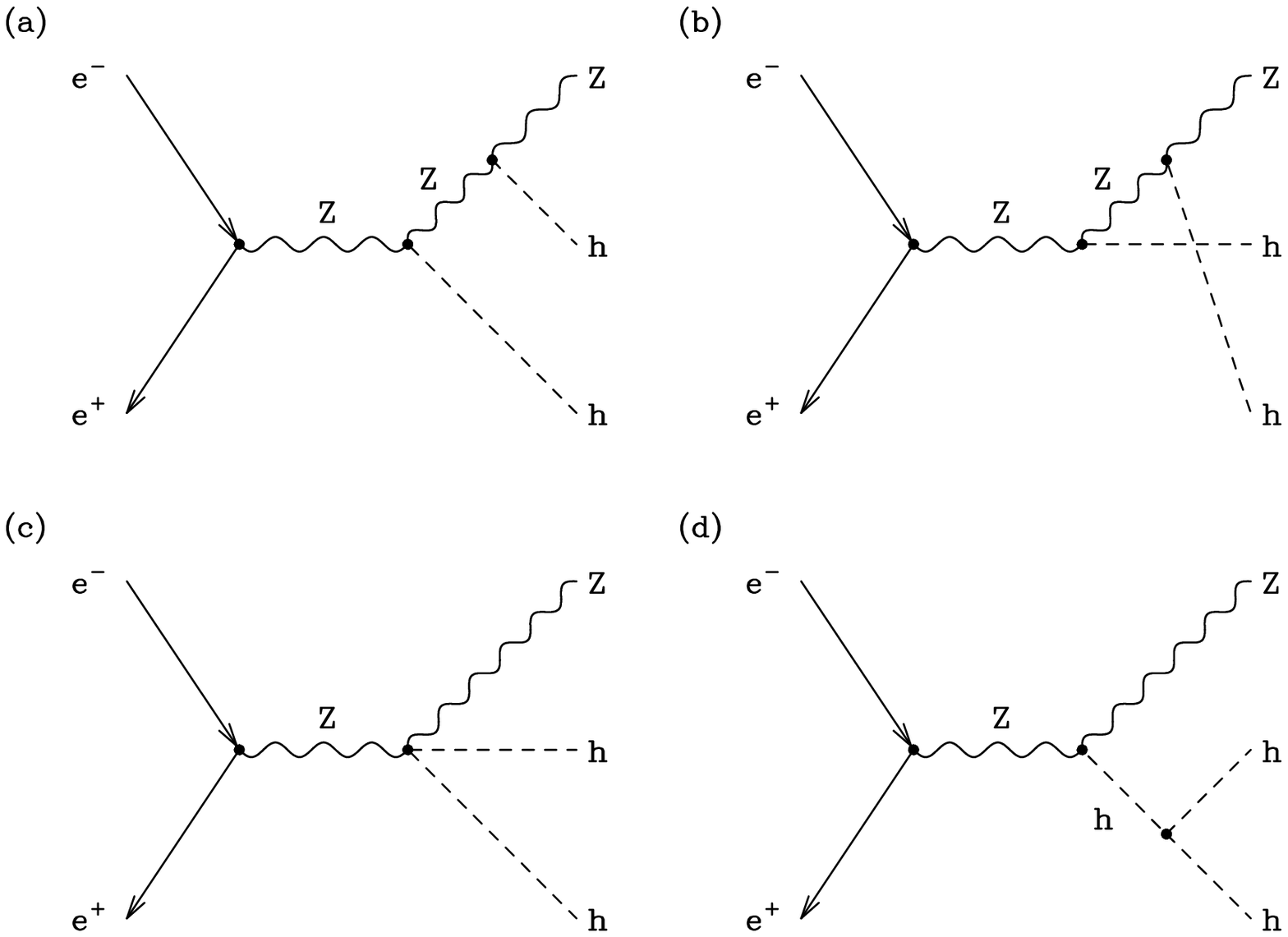}
\vspace*{-5mm}
\caption[Feynman diagrams in SM]{
  The relevant Feynman diagrams for $e^+e^-$ $\rightarrow Zhh$ 
in the unitary gauge in the SM.  
}
\eec
\eef

\section{The Standard Model}

  In the unitary gauge of the SM, at the tree level, there 
are four Feynman diagrams relevant to the process 
$e^+e^- \rightarrow Zhh$ (Fig.~1).  
Three of them come purely from the gauge interactions, 
while the other one has a Higgs-boson self-coupling vertex.  
The Higgs self-coupling constant carries information about the 
Higgs potential, and in the SM it is expressed by the Higgs mass.  
\begin{equation}
  g_{hhh} \ = \ 3 \lambda v \ = \ 3 m_h^2/v \ .  
\end{equation}
Examination of this relation will be the first non-trivial test 
for the shape of the Higgs potential.  

  Unfortunately, it is known that the cross section of this 
precess is not large\cite{GSR,BHP,IPKSK}, 
and we cannot expect a sufficiently 
precise determination of the cross section to study various 
models with anomalous Higgs self-couplings.  
Therefore we focus only on the problem of whether the SM can be 
distinguished from a model without the triple Higgs-boson 
self-coupling (NT).\cite{IPKSK,BH,Jikia} 

\bef{tp}
\bec
\hspace*{1mm}
\epsfxsize=12cm
\epsffile{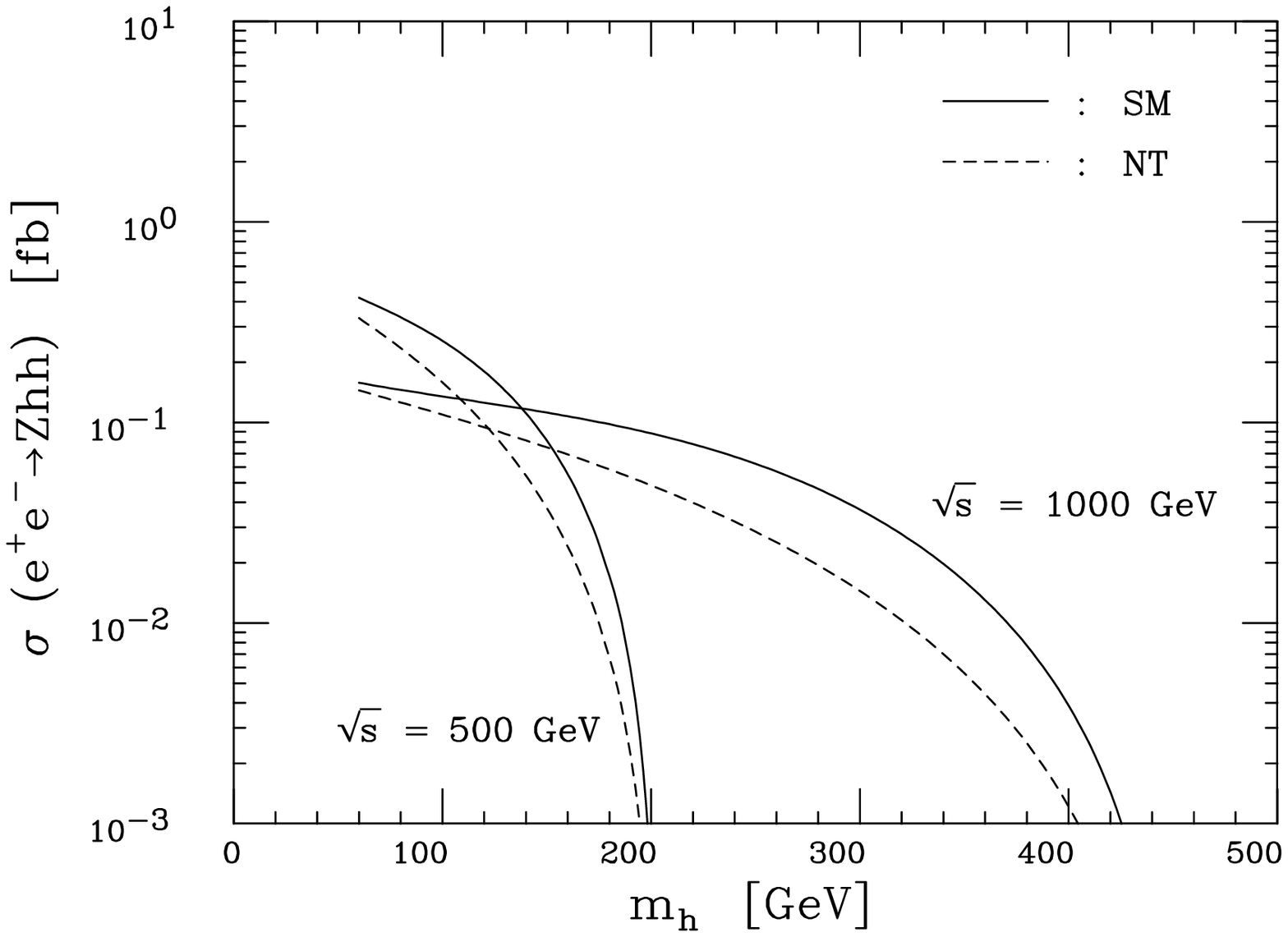}
\vspace*{-5mm}
\caption[Cross section in SM, NHSC]{
  The Higgs-mass dependence of the $e^+e^-$ 
$\rightarrow Zhh$ cross section in the SM (solid) and 
in the NT (dashed), for $\sqrt{s}$~= 500 GeV and 1000 GeV. 
}
\eec
\bec
\hspace*{1mm}
\epsfxsize=11cm
\epsffile{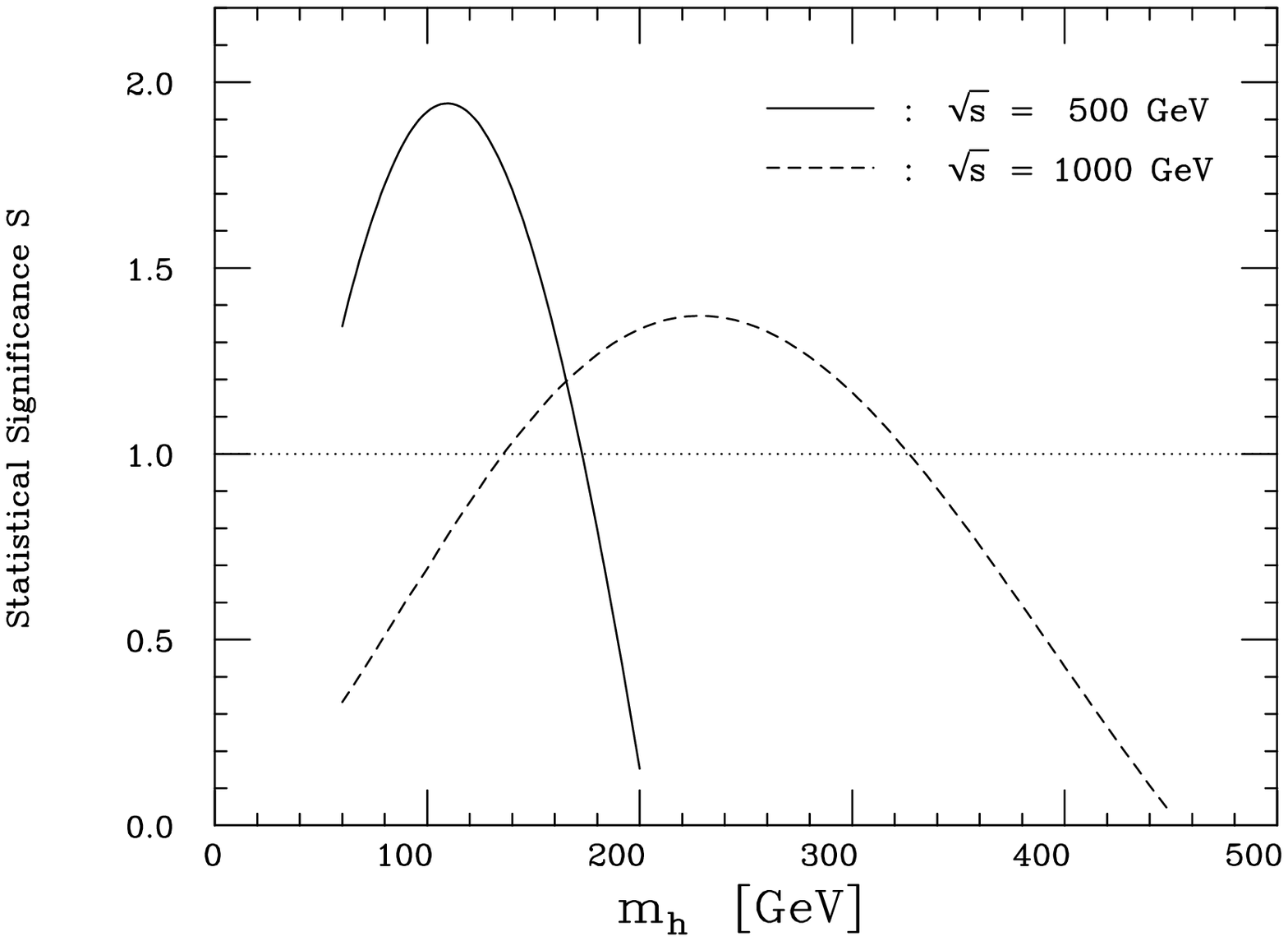}
\vspace*{-9mm}
\caption[Statistical significance]{
  The statistical significance $S$ defined in Eq.(2) 
  with $N_{model}=N_{NT}$ and $N_{obs}=N_{SM}$
  as functions of the 
  Higgs mass for $\sqrt{s}$~= 500 GeV (solid) and 1000 GeV (dashed). 
  We assumed perfect detection efficiency, 100 fb$^{-1}$ of 
  the integrated luminosity and no beam polarization.  
}
\eec
\eef

  The dependence of the cross sections in these models 
on the Higgs-boson mass is plotted in 
Fig.~2 for $\sqrt{s}$ = 500 GeV and 1 TeV.
Here we assume no beam polarization.  
The cross sections are much smaller than 1 fb, and they  quickly drop 
as they approach the kinematic limit.  
We found that the dominant Feynman amplitude in the unitary
gauge is the one with the $ZZhh$ contact coupling (Fig.~1(c)) in the 
most of the parameter space.  
All of the diagrams interfere with each other, but the
amplitude of the diagrams Fig.~1(c) and (d) have always the same
argument.  
The difference between the two models is not so large; however, it is 
notable that the cross section in the SM is always greater than 
that in the NT.  
The ratio of the cross sections of the two models depends on 
the Higgs-boson mass
and the collider energy; it varies from about 0.3 to 0.9.  

  Assuming 100 fb$^{-1}$ of integrated luminosity, 100\%
detection efficiency and no beam polarizations, we can estimate
the statistical significance of the difference between the two models 
as presented in Fig.~3.  
The expected number of the signal events in a model (e.g. SM or NT) 
is $N_{model}$ $= \sigma_{model} \; \times$ 100~fb$^{-1}$, where 
$\sigma_{model}$ is the cross section of $e^+e^-$ $\rightarrow Zhh$ 
process in the model.  
Such a model will be rejected at the 1 $\sigma$ level (stat. only) if 
the statistical significance, 
\begin{equation}
  S = | N_{model} - N_{obs} | / \sqrt{N_{obs}} \ ,  
\end{equation}
exceeds unity, where $N_{obs}$ is the observed number of events, and 
the denominator in R.H.S. represents the statistical fluctuation of 
$N_{obs}$.  
If the SM is correct, we can expect $N_{obs}$ $= N_{SM}$ in 
average, and from Fig.~3, we found that the NT will be
rejected for a Higgs-boson mass of up to 170 GeV with $\sqrt{s}$ 
= 500 GeV, and up to 320 GeV at a 1 TeV collider.  

  The dependence of the cross section on beam polarization is 
simply factored out, since the electron is coupled only to the 
$Z$ boson, as is shown in Fig.~1.  
The ratio of the cross section with both beams polarized to that 
with no polarization is expressed as follows:  
\begin{eqnarray}
       \frac{\sigma(P_{e^-},P_{e^+})}{\sigma(0,0)} 
 & = & (1-P_{e^-} \cdot P_{e^+}) - 
       \frac{1-4 \sin^2 \theta_W}
             {1-4 \sin^2 \theta_W + 8 \sin^4 \theta_W}
       (P_{e^-} - P_{e^+}) \nonumber \\
 & \simeq & (1-P_{e^-} \cdot P_{e^+}) - 0.144 (P_{e^-} - P_{e^+}) \ , 
\label{eq3}
\end{eqnarray}
where $P_{e^-}$ and $P_{e^+}$ are the polarizations of the
electron beam and the positron beam, respectively, and 
$\theta_W$ is the Weinberg angle.  
Highly polarized beams with right-handed electrons ($P_{e^-} \sim +1$) 
and left-handed positrons ($P_{e^+} \sim -1$) are 
preferred to suppress 
background processes such as $e^+e^- \rightarrow 
W^+W^-Z$.  
If the both of the beams can be perfectly polarized, the factor of
Eq.3 is as large as 1.7.  

\bef{t}
\bec
\hspace*{1mm}
\epsfxsize=10cm
\epsffile{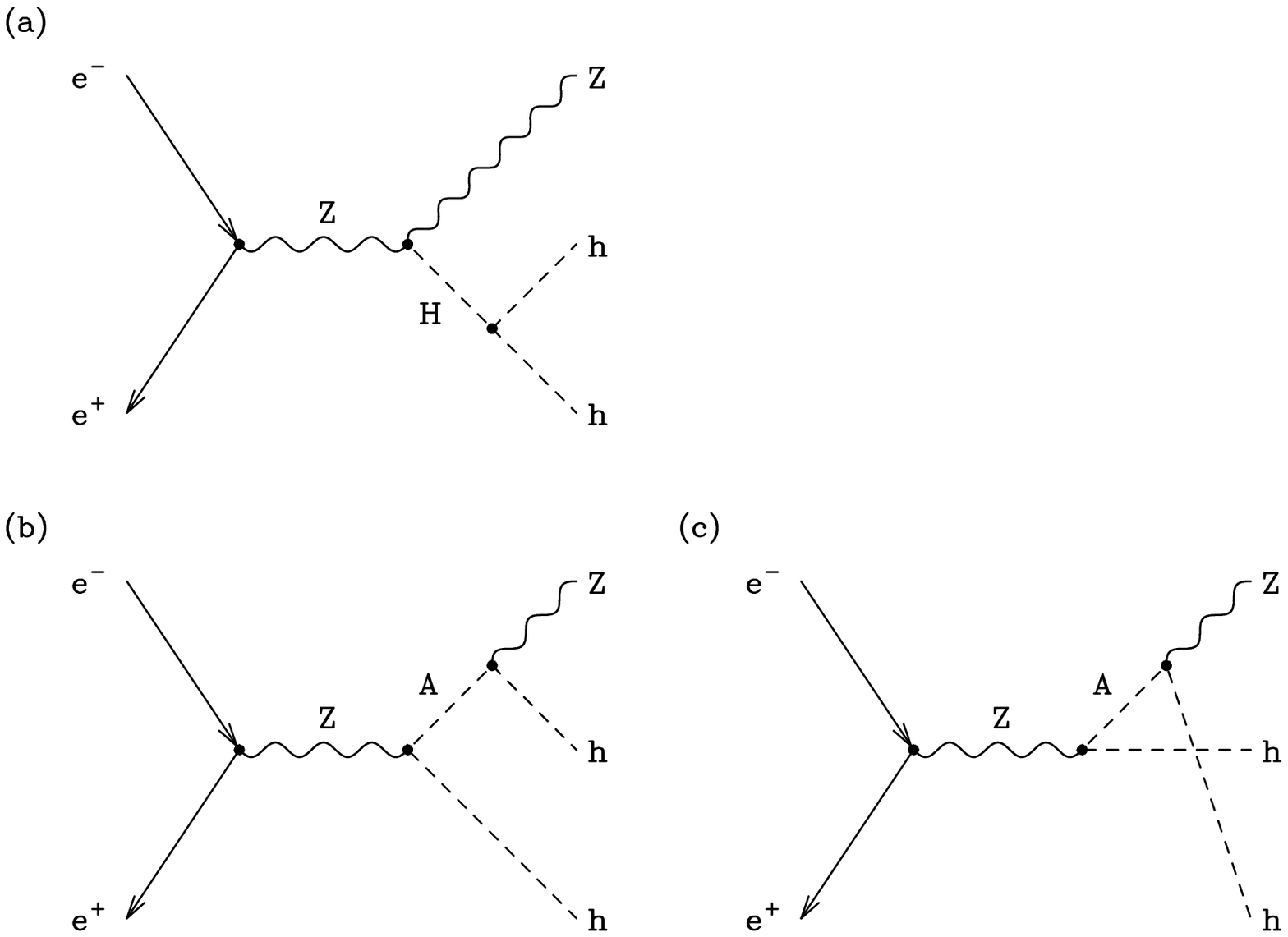}
\vspace*{-8mm}
\caption[Feynman diagrams in MSSM]{
  The additional Feynman diagrams for $e^+e^-$ $\rightarrow Zhh$ 
in the unitary gauge in the MSSM.  
All the relevant diagrams are in this figure and in Fig.~1.  
}
\eec
\eef

\section{The MSSM}

In the Minimal Supersymmetric extension of the Standard Model
(MSSM), there are three additional diagrams (Fig.~4) in addition to 
Fig.~1 in the SM, which are relevant to the process $e^+e^- 
\rightarrow Zhh$, one with the exchange of a heavy Higgs boson ($H$) 
(Fig.~4(a)) and two with the exchange of a pseudo-scalar ($A$) 
(Fig.~4(b,c)).  
(In the MSSM, we denote the light Higgs boson as ``$h$".)  
In some regions of the SUSY parameters and the collider energy, 
the real production of the $H$ and the $A$ and their subsequent decays
through $H \rightarrow hh$ and $A \rightarrow Zh$ are possible,
and in such cases a large cross section is obtained.  

\bef{t}
\bec
\hspace*{1mm}
\epsfxsize=14cm
\epsffile{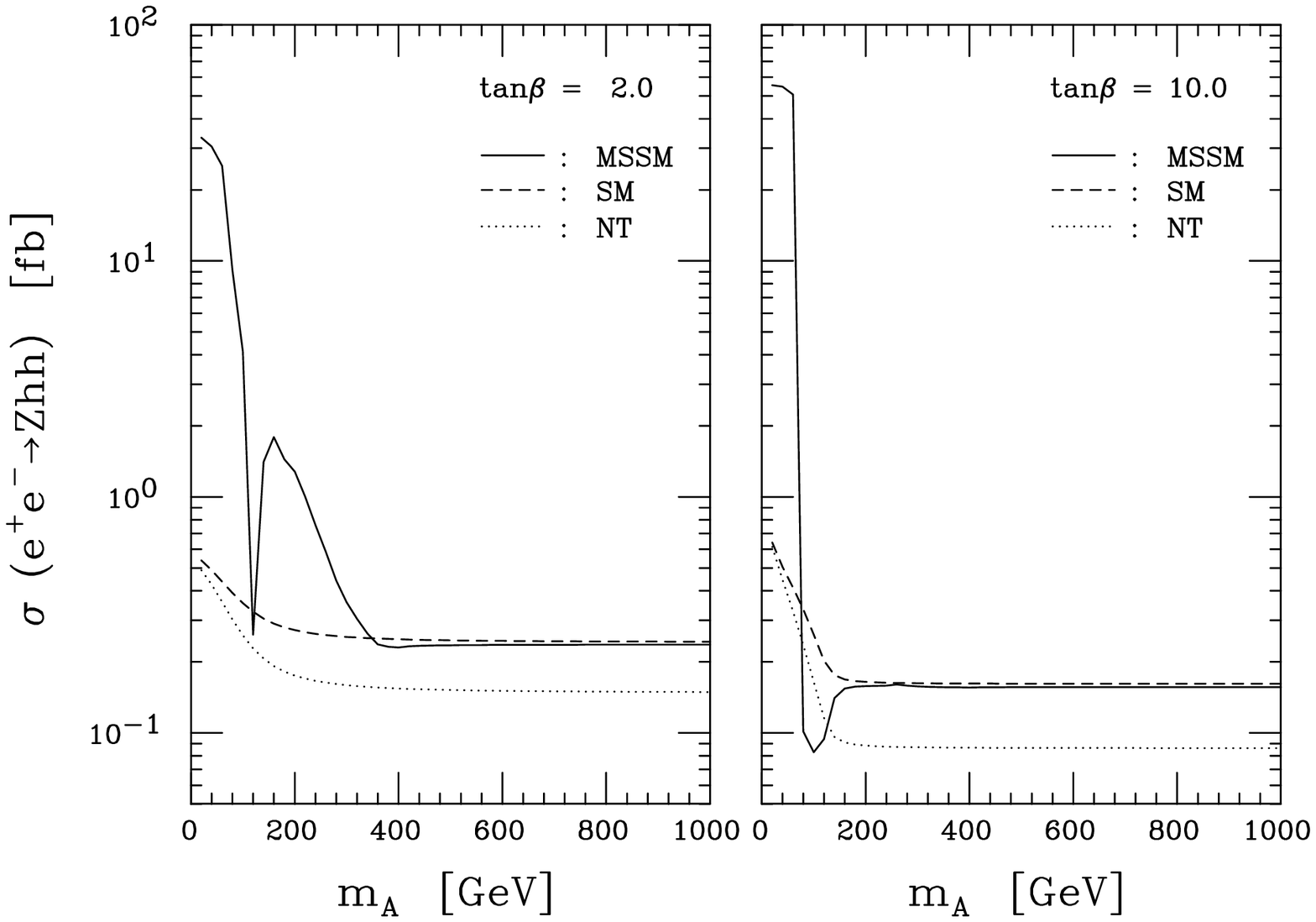}
\vspace*{-5mm}
\caption[Cross section in MSSM: fixed tan(beta)]{
  The $m_A$ dependence of the $e^+e^-$ $\rightarrow Zhh$ 
cross-section in the MSSM (solid) for $\tan \beta$ =~2 and 10 
at $\sqrt{s}$ =~500 GeV. 
We take $m_t$ =~170 GeV and $m_{stop}$ =~1 TeV. 
In this figure the Higgs mass is given as a function of $m_A$. 
For comparison, the SM (dashed) and the NT (dotted) cross-sections 
with the same Higgs mass are also shown.
}
\eec
\eef
\bef{hp}
\bec
\hspace*{1mm}
\epsfxsize=13cm
\epsffile{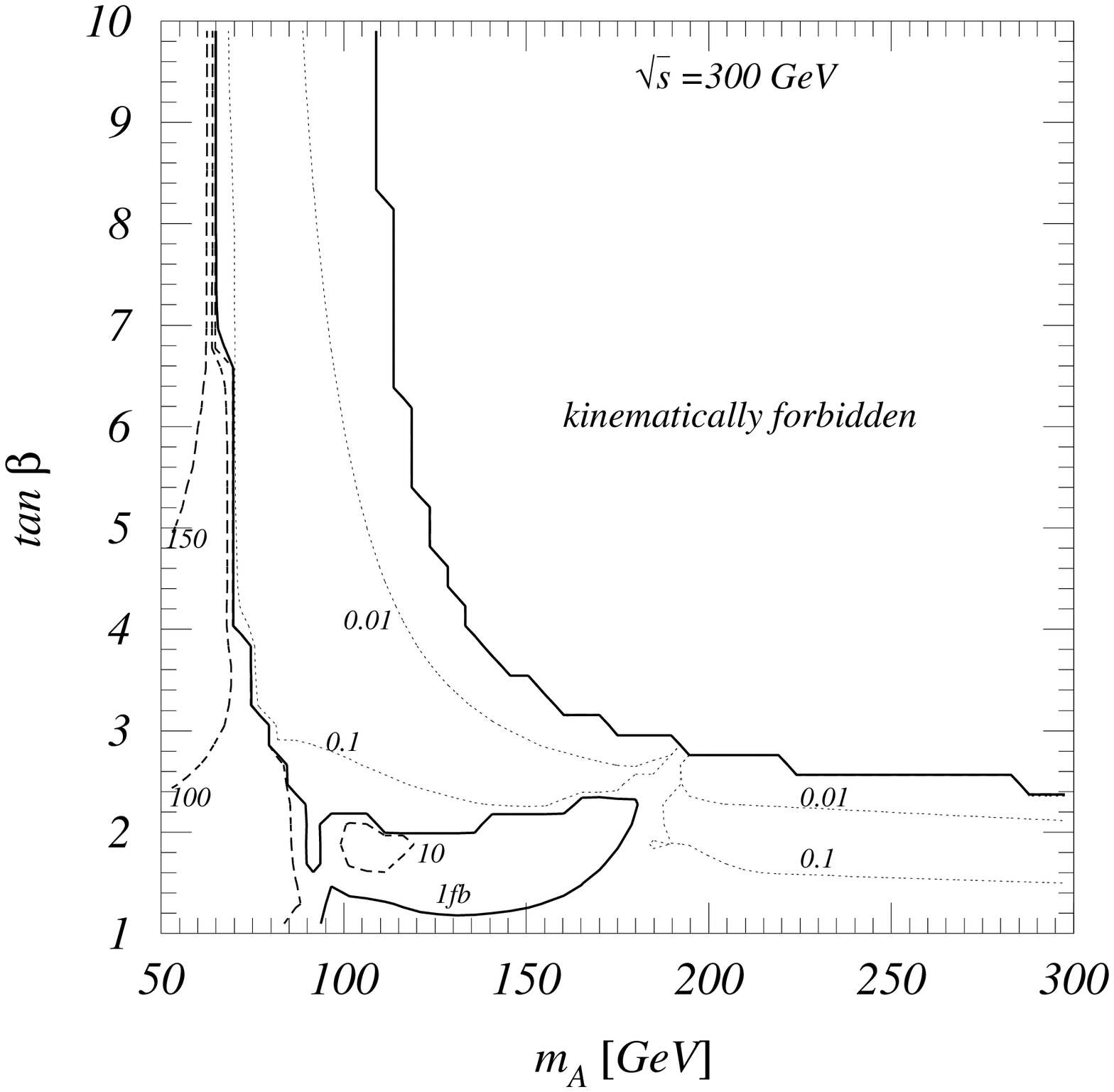}
\vspace{-15mm}
\caption[Cross section in MSSM]{
  The contour plot of cross section of $e^+e^-$ $\rightarrow Zhh$ 
in the MSSM at $\sqrt{s}$ =~300 GeV.  
The top and the stop masses are the same as in Fig.~5.  
}
\eec
\vfil
\eef
\bef{hp}
\bec
\hspace*{1mm}
\epsfxsize=13cm
\epsffile{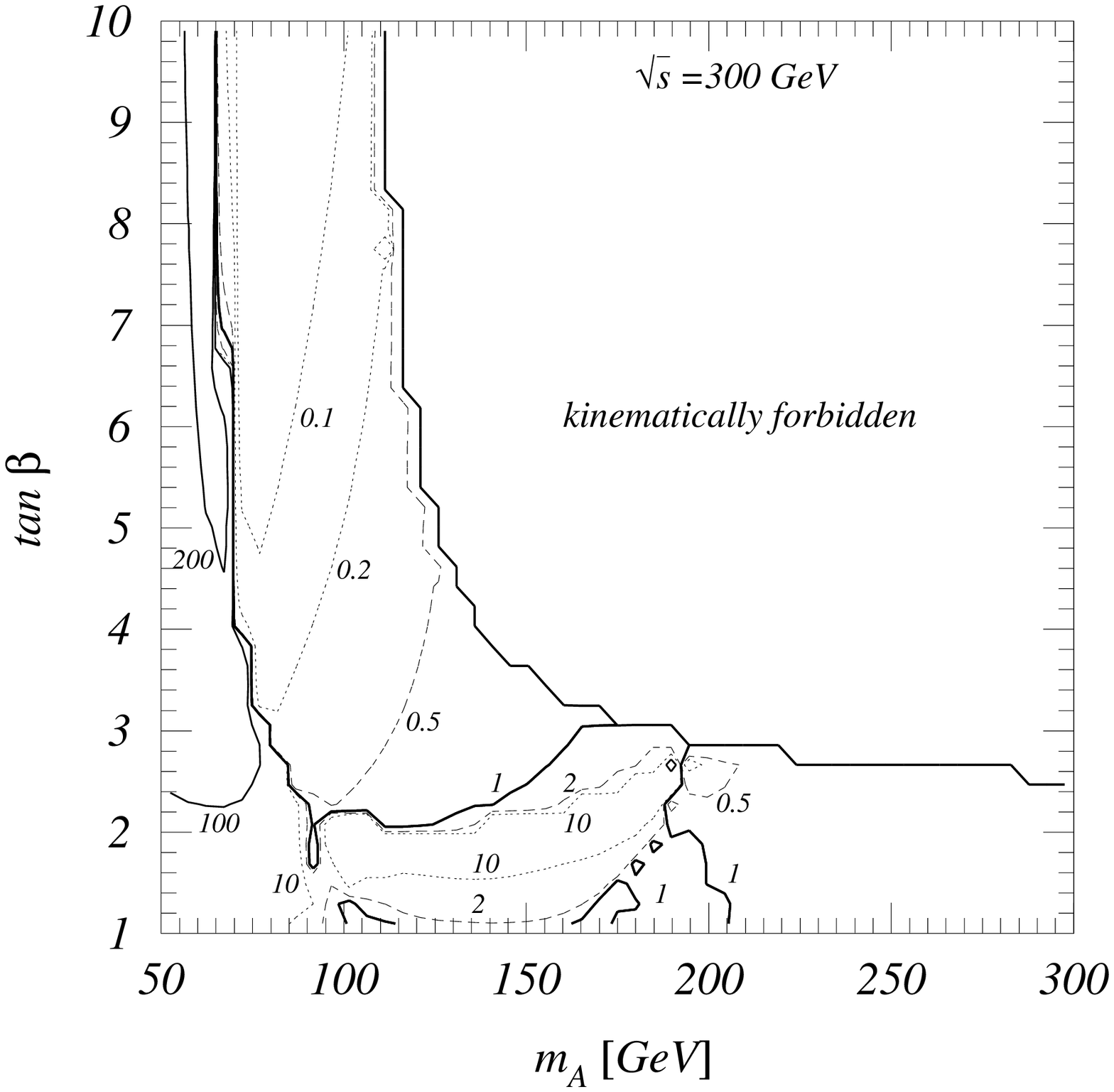}
\vspace{-15mm}
\caption[Cross section ratio in MSSM]{
  The ratio of the $e^+e^- \rightarrow Zhh$ cross-section
in the MSSM to that in the SM.  
The collider energy is 300 GeV, and the Higgs mass in the SM is 
adopted to be the same as $m_h$ in the MSSM at each point in the 
figure.  
The top and the stop masses are the same as in Fig.~5.  
}
\eec
\vfil
\eef

  The dependence on the beam polarization 
of the cross section is precisely the same as in the SM.  

  In the MSSM, once $m_{A}$, $\tan \beta$, the top-quark  mass and the 
top-squark mass~($m_{stop}$) have been specified, the light Higgs 
mass, $m_h$, is predicted. 
We choose $m_t$ =~170 GeV and 
$m_{stop}$ =~1 TeV and plot the MSSM $e^+e^- \rightarrow Zhh$
cross-section versus $m_A$ in Fig.~5. 
In this figure, the light Higgs mass $m_h$ is given as a
function of $m_A$.  
For comparison we also plot the SM and NT cross-sections with
the same $m_h$ as in the MSSM.  
Large cross section can arise in the MSSM in the region where the
poles of the heavy Higgs and/or the pseudo-scalar are hit.  
This is especially remarkable for the heavy Higgs pole.  
However, the cross section rapidly approaches to the SM value as
$m_A$ goes large.  
In this limit the lightest Higgs boson behaves just like the SM 
Higgs boson.

The contour plots of the cross section in the plane of $m_A$ and
$\tan \beta$ in the MSSM with $\sqrt{s}$ = 300 GeV can be found in 
Fig.~6.  
The lightest Higgs mass $m_h$ varies from 48 GeV to 124 GeV in
this figure.  
The process $e^+e^- \rightarrow Zhh$ is
kinematically forbidden in the upper right region.  
In the left-end region of the figure, the cross section is
as remarkably large as 100 fb, and there is also a horizontal area
at the bottom of the figure 
where the cross section exceeds to 10 fb. 
Both areas of large cross section correspond to the pole of the
heavy Higgs.

Fig.~7. shows the ratio of the cross sections of the MSSM and 
the SM.  
Again, the resonant regions correspond to a large ratio.  
It is quite interesting that there is an area with a cross-section 
ratio sizably smaller than unity at around $m_A$ =~70 -- 130 GeV and 
$\tan \beta$ $>$~2.5.  
In this region, no internal Higgs boson is on-shell, and the
$ZZh$ coupling is suppressed.

\section{Conclusions}

  The process $e^+e^- \rightarrow Zhh$ is important to examine
the Higgs potential.  

  We found that the SM and the NT can be distinguished at 
the level of one standard deviation for 
a rather light Higgs mass with 100 fb$^{-1}$ integrated luminosity.  

  In the MSSM case, the cross section is enhanced if the $H$ 
or $A$  pole is accessible.  
In the case where no pole is accessible, the cross section is 
significantly suppressed relative to the SM at around $m_A$ =~70 -- 
130 GeV and $\tan \beta$ $>$~2.5.  
In this region the MSSM can be 
distinguished from the SM, if enough luminosity is 
available.  

  To perform these studies at a future linear collider, a
high luminosity, a high efficiency and high beam polarizations are
required.  

\section*{Acknowledgements}

  The authors would like to thank Rob Szalapski
  for reading the manuscript and useful comments. 	
  The present work was partially performed while 
  one of the authors (I.W.) was at Fermilab and
   supported by
  `Japan/U.S. Cooperation in the Field of High Energy Physics'.


\end{document}